# Heuristics for the Variable Sized Bin Packing Problem Using a Hybrid P System and CUDA Architecture

QADHA'A ALENEZI*, HOSAM ABOELFOTOH**, BADER ALBDAIWI***, AND MOHAMMAD ALI ALMULLA****

*Computer Science Department, Kuwait University, Kuwait, e-mail: qadha@live.com
**Computer Science Department, Kuwait University, Kuwait, e-mail: hosam@cs.ku.edu.kw
***Computer Science Department, Kuwait University, Kuwait, e-mail: bdaiwi@cs.ku.edu.kw
****Computer Science Department, Kuwait University, Kuwait, e-mail: almulla@cs.ku.edu.kw

## ABSTRACT

The Variable Sized Bin Packing Problem has a wide range of application areas including packing, scheduling, and manufacturing. Given a list of items and variable sized bin types, the objective is to minimize the total size of the used bins. This problem is known to be *NP*-hard. In this article, we present two new heuristics for solving the problem using a new variation of P systems with active membranes, which we call a *hybrid P system*, implemented in CUDA. Our hybrid P-system model allows using the polarity and labels of membranes to represent object properties which results in reducing the complexity of implementing the P-system. We examine the performance of the two heuristics, and compare the results with those of other known algorithms. The numerical results[1] show that good solutions for large instances (10000 items) of this problem could be obtained in a very short time (seconds) using our CUDA simulator.

## 1. INTRODUCTION

The Variable Sized Bin Packing Problem (***VSBPP***) is an NP hard problem. It contains the classical Bin Packing Problem (***BPP***), as the BPP has only one type of bins. There are many applications in practical life for the VSBPP. For example, in the truck loading problem, we use trucks with different sizes to load items. We can use as many trucks of each size as we want. The goal is to minimize the total capacity (or cost) of selected trucks. Also, the machine-scheduling problem can be considered as a VSBPP. This problem appears when different classes of processors are used to parallelize the processing of a set of processes with a known processing time. The goal is to minimize the total cost of the associated processors (Coreia, I., Goveia, L. and da Gamma, F., 2008). Transmitting requests through the network is another application of the VSBPP. It is assumed that a set of unit-time tasks is to be transmitted through a network, each task requiring a specific network capacity. The network

---

provides different capacities and here the goal is to schedule the tasks in such a way to minimize the total used capacity of the network (ZHANG, 2002).

**Formal Definition of VSBPP**

Formally we can define the VSBPP as follows. We are given $n$ types of bins. Let $B_i$, $1 \leq i \leq n$ denote the capacity of each bin of type $i$, such that $B_1 > B_2 > ... > B_n > 0$. The VSBPP is to find a packing of a list of non-divisible items of different weights ($w_1, w_2 ... w_j ..., w_m$), where, $m$ is the number of items in the list, $w_j$ is the weight of item $j$, $1 \leq j \leq m$ into a set of bins such that the total capacity of used bins is minimum. We assume that for each type of bin, there are an infinite number (unlimited supply) of bins. The weight of the largest item is less than or equal to the capacity of the largest bin type (ZHANG, 2002).

The VSBPP contains the classical one-dimensional BPP as a special case. In the BPP all the bins have the same capacity and cost ($n = 1$). The BPP is an NP-hard problem and thus, the VSBPP is NP-hard as well" (Coreia, I., Goveia, L. and da Gamma, F., 2008).

Most recognized packing heuristics for BPP are *Best Fit* (*BF*), *Worst Fit* (*WF*), *and First Fit* (*FF*). The BF keeps only a specific number of bins open (*active*) until the end of the packing process. The heuristic looks for a suitable bin to pack the new item. The new item will be eligible to be packed in an open bin with the least room left over. If there is more than one bin satisfying this criterion, the item will be packed in the bin with the lowest index. The WF is similar to the BF heuristic. The only difference between them is packing the new item in the bin with the most room left. In the FF the new item will be packed in the first active bin -with the lowest index- which has enough space to pack the item. If the current item cannot fit in any of the active bins, a new bin will be opened to pack that item. The bin will be closed only if it is completely filled with items (Malkevitch, "Bin Packing", American Mathematical Society, 2004). For these heuristics to be applied to VSBPP, one must have additional heuristics for choosing amongst the different types of bins at each step.

In this paper, we present two new heuristics for the VSBPP based on a P system and implemented in CUDA. We use a new implementation of P systems which we call a *hybrid* P system. Section 2 is an introduction to P systems and related concepts. Section 3 gives more details on P systems with Active Membranes. In Section 4, we give an overview of General-Purpose Computing on Graphics Processing Units (GPGPU) and Nvidia Compute Unified Device Architecture (CUDA). Section 5 includes previous work related to VSBPP and cites the work done on simulating P systems using GPGPU and CUDA. Section 6 presents our two new heuristics. Section 7 contains the numerical results. Section 8 is conclusion and future work.

## 2. MEMBRANE COMPUTING – P SYSTEM

Membrane computing was first introduced by Gheorghe Păun on 1998 (Paun, P Systems with Active Membranes: Attacking NP Complete Problems, May 1999). Membrane computing is a branch of Natural computing which refers to any computational model that is based on an inspiration from nature (Castro, 2006). This includes Cellular Automata, Neural

Computation, Evolutionary Computation, Swarm Intelligence, Artificial Immune System, Amorphous Computing, etc. (Kari, L. and Rozenbergrz, G., October 2008).

Membrane computing is an abstract computing model based on simulating the structure and the functions of living cells and the way the cells are organized in tissues. Therefore, Membrane Computing is a distributed and parallel computing model. The initial membrane systems are interlaced in a hierarchical structure of **compartments** similar to the cells or regions surrounded by **membranes**. Each membrane region consists of multi-sets of **objects**, **evolution rules**, **communication rules**, and **transformation rules**. Various additional features and variants have been added to the initial membrane systems; therefore new classes of Membrane computing are invented. All these classes of Membrane systems are generally referred to as **P systems**, relative to the name of the founder of this computing model Gheorghe Păun. (Paun, G., and Rozenberg, G., 2009).

Many of the hard problems could be solved (i.e. optimal or near-optimal solution could be obtained efficiently) by enhanced classes of P systems (Paun, Introduction of Memebrane Computing, 2006) (Paun, G., and Rozenberg, G., 2009). P systems computational models have several features that are interesting for many applications. It is easy to be programmed, scalable/extensible, distributed, parallel, non-deterministic, transparent, and communicable. These features make the P system a suitable model for dealing with NP-hard problems with exponential search space (Paun, P Systems with Active Membranes: Attacking NP Complete Problems, May 1999).

The main three variants of P systems are cell-like, tissue-like, and neural-like. In the cell-like class the main component is the membrane structure, the membranes are arranged hierarchically. The objects are placed in compartments that are predetermined. The objects are denoted by specific symbols. There are several forms of rules; each one has a specific role. Generally the tissue-like class consists of one membrane cell in a common environment. Multi-sets of objects are placed in the cells, and the common environment. The cell can communicate directly or through the environment using the variant types of the rules. The neural-like class is similar to the tissue-like class in that the cells are one membrane cell. The cells are placed arbitrary on the nodes of a graph, and there are multi-sets of objects in each cell. In the neural-like P systems the cells have states that govern the evolution (Paun, Introduction of Memebrane Computing, 2006) (Paun, G., and Rozenberg, G., 2009) (Ibarra, O. and Paun, G., 2006-2007). The cell-like class was the first class invented in this branch of Natural Computing. The heuristics proposed in this paper mainly uses the cell-like P system with active membranes as a framework.

## 3. P SYSTEM WITH ACTIVE MEMBRANE

P system with active membrane is one of the variants of P systems. In this variant, the membranes can evolve by changing their properties or numbers. The number of membranes could be decreased by *dissolution* operation or increased by the *division* operation. Since P system is a parallel computing system, the number of membranes can be increased exponentially in linear number of steps. For example, by dividing a membrane by *n* operations we can get $2^n$ copies of that membrane. This helps creating an exponential working space for NP-complete problems (Paun, Introduction of Memebrane Computing, 2006).

P system with active membranes can be defined formally as: $\pi$ = (*O, H, e, $\mu$, $m_1$, $m_2$, ... , $m_p$, R*) , where *m ≥ 1:* is the initial degree of the P system, *O:* the alphabets that represents the objects, *H:* finite set of membrane labels, *e:* finite set of signs representing the **polarity** of the membrane {*-, +, 0*}, *$\mu$:* membrane structure consists of a number (*p*) of membranes that are initially with **neutral polarity 0**, each membrane is labeled by an element of *H*. For example [ [ [ ]$_2$ ]$_1$ [ ]$_3$ ]$_0$ represents an *outer* membrane labeled *0* (usually called the *skin)* that contains two internal membranes labeled *1* and *3*. Membrane *1 contains an internal* membrane labeled *2*. The simplest membrane structure is: [ ]$_h^e$, where, *h* is the label of the membrane and *e* is the membrane polarity. $m_1, m_2, ... , m_p$ : strings of alphabets over *O*, representing the multisets of objects in the membrane structure. *R:* Finite set of development rules associated with the regions of the P system. There are five types of the development rules (we give an example of each type):

**A. Object Evolution Rules:** This type of rules is evolving only the objects -not the membranes- and it depends on the polarity and labels of the membranes.

$$[ \ a \rightarrow v ]_h^e, \text{ where, } h \in H, e \in \{+, -, 0\}, a \in O, v \in O^*$$

The element *a* will be evolved to *v* if the polarity of the membrane *h* is *e*.

**B. In-Communication Rules:** In this type of rules an object is injected **into** the membrane and it could be evolved during this process for another object. The polarity of the membrane could be changed but not the label of the membrane.

$$a[ \ ]_h^{e1} \rightarrow [ \ b ]_h^{e2}, \text{ where, } h \in H, e1, e2 \in \{+, -, 0\}, a, b \in O$$

The object *a* is injected into the membrane *h*, during the transformation process the object *a* evolved to *b*. The polarity of the membrane *h* changed from *$e_1$* to *$e_2$*.

**C. Out-Communication Rules:** In this type of rules an object is sent **out** the membrane and it could be evolved during this process for another object. The polarity of the membrane could be changed but not the label of the membrane.

$$[ \ a ]_h^{e1} \rightarrow [ \ ]_h^{e2} b, \text{ where, } \boldsymbol{h} \in H, \boldsymbol{e1, e2} \in \{+, -, 0\}, \boldsymbol{a, b} \in O$$

The object *a* is sent outside the membrane *h*, during the transformation process the object *a* evolved to *b*. The polarity of the membrane *h* changed from *$e_1$* to *$e_2$*. Generally the label of the membrane could be changed in the (**in** and **out**) communication rules.

**D. Dissolving Rules:** As a result of interaction of the objects the membrane could be dissolved and the objects modified.

$$[ \ a ]_h^e \rightarrow b, \text{ where, } \boldsymbol{h} \in H, \boldsymbol{e} \in \{+, -, 0\}, \boldsymbol{a, b} \in O$$

The membrane *h* is dissolved and the object *a* evolved to *b* and dumped to the parent membrane.

**E. Division Rules:** In this rule as a result of object interaction the membrane will be divided into two membranes. The new membranes could be with different polarities and different labels. Some objects will evolve for new objects; the lasting components -objects or inner membranes- will be duplicated in all the new membranes.

$$[ \ \ a]_{h1}^{e1} \ \to [ \ \ b]_{h2}^{e2}[ \ \ c]_{h3}^{e3}, \ where \ \textbf{\textit{h1, h2, h3}} \in H, \textbf{\textit{e1, e2, e3}} \in \{+,-,0\}, \textbf{\textit{a, b, c}} \in O$$

The membrane $\textbf{\textit{h}}_{\textbf{\textit{1}}}$ will be divided into two membranes one of label $\textbf{\textit{h}}_{\textbf{\textit{2}}}$ and one of label $\textbf{\textit{h}}_{\textbf{\textit{3}}}$. The new membranes could be with different or the same polarity of the membrane $\textbf{\textit{h}}_{\textbf{\textit{1}}}$. The object $\textbf{\textit{a}}$ will evolve to new objects $\textbf{\textit{b}}$ and $\textbf{\textit{c}}$.

Using the P system as a computational model requires testing for the applicability of each rule and choosing randomly between those eligible for application. In parallel execution of rules, care must be taken when more than one rule are applicable to the same object.

The P system model is a parallel model that can exploit the new technology of GPGPU to allow for developing new heuristics for NP-hard problems.

## 4. GPGPU and CUDA

### 4.1. GPGPU

Graphics Processing Units (GPUs) are massively parallel processors. They are accelerating the graphics computation by taking the job from the CPUs. Today's GPU is extremely fast because of its high parallelism. GPGPU is the use of GPUs to solve general computational problems (other than Graphics). In the recent years, a large interest in the GPGPU approach has appeared aiming at obtaining near-optimal or approximations to NP-Complete problems in a relatively short time. The start of GPGPU was in 2002 and its evolution was slow until 2007 when NVIDIA released CUDA (McClanahan, History and Evolution of GPU Architecture, 2011) (Owens, J., Luebke,D., Govindaraju N., Harris, M., Krüger,J., Lefohn, A., and Purcell, T., August 2005), (NVIDIA CUDA, Programming Guide, 2007). When CUDA was released, the approach of GPGPU grew rapidly. Still the GPGPU is growing with the improvement of NVIDIA's products (Zibula, 2010).

### 4.2. CUDA

CUDA is a hardware and software architecture that was developed by NVIDIA in 2007 to manage the GPGPU and support heterogeneous computation. With CUDA, the serial part will be executed by the CPU and parallel parts by the GPU. CUDA allowed the developer to use C language, Fortran, OpenCL, Java ... etc. This feature makes GPU computation easier even for the novice programmer. CUDA succeeded in qualifying the GPU for the non-graphical operations by overcoming the difficulties that were facing this approach. The GPGPU computation has made a significant progress after the emergence of CUDA.

- **CUDA Programming Model:** With CUDA, the GPU operates as a coprocessor for the CPU. Programming through CUDA makes the CPU perform as a *host* and the GPU performs as a *compute device* with the capability of executing a large number of *threads* in parallel. The portion of the code that needs to be executed in parallel with independent data will be executed on the GPU device on several different threads. This piece of code is called **kernel**, it is compiled to the device instruction set then loaded on the device with all required data. The *kernel call* has the following format: *Thread_code_name<<<number_of_blocks,number_of_threads >>> (parameter_list)*.

There are independent DRAMs for both host and device; however the data can be copied back-and-forth between host memory and device memory.

A thread is a set of instructions that are to be processed by GPU (device). Each thread has two indices (block index and thread index). The set of threads executing the same kernel are organized in a *grid*. The grid is composed of a number of *thread blocks*. Thread block is a batch of threads to be executed in parallel that can cooperate together by sharing data through a *fast shared memory*. A grid of thread blocks is a batch of thread blocks of the same dimensions, sizes, and they are executing the same kernel. Threads can synchronize their execution by suspending the threads of a block until all of them reach the synchronization point. Threads that are in different blocks of the same grid cannot communicate or synchronize with each other. This hierarchy allows the kernel to be compiled only once for each device.

- **CUDA Memory Model:** There are various types of memory spaces available on a GPU device, these memory spaces are classified as: The memory space that can be accessed by a thread is thread's registers (32-bit) and thread's local memory (parallel data cache). The memory space that can be accessed by all threads of a thread block is the block's shared memory. The memory space that can be accessed by all thread blocks of a grid are grid's *global* memory, grid's *constant* memory (read only), and grid's *texture* memory (read only).

- **CUDA Hardware Implementation:** The GPU is built of a set of multithreaded, multicore, streaming multiprocessors. The multiprocessors consist of *Single Instruction Multiple Thread architecture (SIMT)*. SIMT means that each processor in the multiprocessor executes the same instruction with different threads (and subsequently different data). Each multiprocessor in the GPU has a (read and write) shared memory accessed by its processors only. A read-only texture cache and constant cache are shared by all processors of the multiprocessor. A set of 32-bit registers is assigned to each processor of the multiprocessor.

- **CUDA Execution Model:** A CUDA program is executed on the CPU (the host), as soon as it reaches a call for a kernel on the GPU, it will invoke that kernel on the GPU. The kernel will be executed by a grid of thread blocks. Each multiprocessor will handle one or two blocks using time slicing. Each block will be divided into SIMT sets of threads (Warps). Each warp has the same number of the threads. The warp is executed by a multiprocessor. A thread scheduler will swap from one warp to another to maximize the performance of the multiprocessor.

A block can be executed only on one multiprocessor. More than one block can be executed on the same multiprocessor. The order of warp execution within a block is undefined but it can be synchronized to arrange the accessing of the shared and global memory. The order of the thread blocks execution is undefined and it cannot be synchronized, because of that threads from different blocks cannot communicate to keep the shared and global memory safe from wrong updates (NVIDIA CUDA, Programming Guide, 2007).

## 5. PREVIOUS WORK

On 2003, Kang and Park introduced two greedy algorithms (Kang, J. and Park, S., 2003). On 2002, Guochuan Zhang proposed a number of approximation algorithms to solve online VSBPP (ZHANG, 2002). Haouari and Serairi proposed six optimization-based heuristics for the VSBPP (Haouari, M., and Serairi, M., 2009). Coreia, Goveia, and da Gamma studied the use of a discretized function for solving the VSBPP (Coreia, I., Goveia, L. and da Gamma, F., 2008). Crainic, perboli, Rei, and Tadie proposed a series of lower bounds then they computed the means to measure the quality of their solution. They used these bound to build their heuristics for the problem (Crainic, T., perboli, G., Rei, W., and Tadie, R., 2010). Later on 2011 the authors introduced Lower bounds for the Generalized BPP. They presented two formulations for the problem, an aggregate knapsack lower bound, and column generation-based lower bound (Baldi, M., Crainic, T., Perpoli, G. and Tadei, R., 2011). On 2012 they proposed two methods to solve the problem, an exact method using the branch-and-price search and a heuristic using the beam search (Baldi, M., Crainic, T., Perpoli, G. and Tadei, R., 2012). On 2014 they applied a worst case analysis to the Generalized BPP of FFD, and BFD heuristics. Also they proposed two semi-online algorithms to tackle the problem (Baldi, M., Crainic, T., Perpoli, G., and Tadei R., 2014).

As for the P system, Zandron , Ferretti and Mauri published a paper in 2001. They showed that the P systems with active membrane which use one type of membrane division: division of elementary membrane can be used to solve SAT and Hamiltonian Problems in linear time with respect to the input length (Zandron, C., Ferretti, C. and Mauri, G., 2001). On 2004 Perez-Jimenez and Romero-Campero published a paper that provides an effective solution of the standard BPP. They used a family of recognizer P system with active membranes to solve an instance of the BPP (Pérez-Jiménez, M., and Romero-Campero F., 2004). Linqiang Pan, and Carlos Martín-Vide solved the multidimensional 0-1 Knapsack Problem by an algorithm based on the recognizer P system with input and active membrane (Pan, L. and Martin-Vide, C., December 2005). Alhazov, and Perez-Jimenez tried to form a uniform family of P system to solve the Satisfiability of a Quantified Boolean formula (QSAT), which is a PSPACE-complete Problem in their paper, which was published in 2007 (Alhazov A. and Pérez-jiménez, M., 2007). In 2010, Chun Lu, and Xingyi Zhang provided a solution for the Vertex Cover Problem using the Tissue-like P system with cell separation (Lu, C. and Zhang, X., 2010). Aman, and Ciobanu published a paper in 2011 where they provide semi-uniform polynomial solutions for some weak NP-complete problems such as Knapsack, Subset Sum, and Partition Problems (Aman, B. and Ciobanu, G., December 2011).

The research on simulating P systems with GPGPUs has recently attracted many researchers. (Cecilia & etal., 2010) , (Martinez-del-Amor & etal., 2013), (Martinez-del-Amor & etal, Simulating P systems on GPU devices: a Survey, 2015) and (Maroosi & etal, 2014). To our knowledge, this paper is the first to propose a solution (heuristics) to the VSBPP based on P systems and CUDA.

## 6.  A NEW HYBRID P SYSTEM FOR SOLVING THE VSBPP USING CUDA

In this paper we propose two P system-based heuristics to find a solution for the VSBPP using CUDA. However, we define a new type of P systems which we call a *hybrid P-system*. A hybrid P-system differs from the P-system in two ways. First, it extends the set of polarity symbols and labels to represent dynamic properties of the objects as well as the membranes. This feature allows for using the labels to represent the item weight instead of representing the object weight using the multiplicity of the object as defined in the original definition of P-systems. For example, in P systems an item $s_k$ of weight three is represented in the initial configuration as the multiset { $s_k, s_k, s_k$ } (Pérez-Jiménez, M., and Romero-Campero F., 2004). Second, the model allows for selecting an object within a membrane based on a selection criterion (FF, WF, and BF). The selection criterion is represented as a component of an item object label within the membrane. Similar to the recognizer P system model as defined in (Pérez-Jiménez, M., and Romero-Campero F., 2004) two distinguished objects {*Yes, No*} are used to signal a termination condition in the computation.

### 6.1. Problem Description

Given a list of indivisible items of different weights *list* = ($w_1$, ..., $w_m$), where $w_i$ is the weight of the *ith* item, $1 \le i < m$, and $m$ is the number of the items in the list, and $n$ types of bins, $B_i$, $1 \le i \le n$, where $B_i$ denotes the capacity of each bin of type $i$, find a packing of the items such that the total capacity of used bins is minimum.

We assume that for each type of bin, there is an infinite number (unlimited supply) of bins. We also assume that $B_1 > B_2 > ... > B_n > 0$, and the maximum weight of an item is less than or equal $B_1$. Both heuristics perform an initialization step on the host (CPU) then call a kernel to perform the item packing. Let $nb$ and $nt$ be the number of CUDA blocks and threads respectively. The choice of $nb$ and $nt$ will be illustrated in Section 7.

### 6.2. Initialization Step:

This step is common to both heuristics and performed by the host (main):

- *Create the initial configuration of the P system ( $\Pi$ ) as follows*:

    $\Pi = (O, H, e, \mu, m_0, m_1, ... , m_p, R)$

    Objects: O = { $w_1$, ..., $w_m$ } $\cup$ { *FF, BF, WF* } $\cup$ {*Yes, No*} }

Membrane labels: $H = \{ S_1, \ldots, S_l, (b,x,B_1), \ldots, (b,x,B_n) \}$, $x > 0$.

The first set of labels $S$ denote $l$ sublists of items, and $(b,x,B_y)$ denotes a bin number $x$ with capacity (size) $B_y$. $l = nb \times nt$. The maximum sublist size $s = m/l$.

*Polarity:* $e = \{ 0, \ldots, n \} \cup \{ 0 \ldots s \}$ .

Membrane structure: $\boldsymbol{\mu} = \; [ \; []_{S1}^0 \; []_{S2}^0 \; \cdots \; []_{Sl}^0 \; \cdots []_{b,1,B1}^0 \; []_{b,1,B2}^0 \; \cdots \; []_{b,1,Bn}^0 \; ]_0^0$ .

/* The skin membrane contains $l$ membranes (for subsets of items) and one membrane for each type of bin. */

- Initialize skin membrane objects: $m_0 = \{ w_1, \ldots, w_m \}$./* Skin membrane contains all items. */

- Initialize internal membranes: $m_1, \ldots, m_{p} = \Phi$.

- *Rule 1*: $w_i \; [ \ldots ]_{Sl}^k \quad \rightarrow \quad [ \ldots w_i \; ]_{Sl}^{k+1}$ , $k < s$ . This (In-Communication) rule injects item objects into membranes $S1$ , …. $Sl$ . Polarity is incremented.

## 6.3. First Heuristic

1. Distribute the items randomly among the $l$ subsets; until each subset contains $s$ items (Rule 1). This step is executed by the host code (or an initialization kernel).
2. Copy the P system parameters from host memory to device memory.
3. Call the kernel *Pack_1 $<<< nb, nt >>>$* ( *P system parameters*).
4. //Let $bx$ be the block index, $tx$ be the thread index.

   ***Pack_1*** has the following steps: (executed by each thread ($bx, tx$) )

   - Compute subset index $I = bx * nt + tx$. //Thread($bx, tx$) will pack subset $SI$.

   - *Rule 2*: $[]_{b,1,Bi}^0 \rightarrow []_{b,1,Bi,I}^0$ , $i \in \{1 ..n\}$ // Create thread own bins

   - Create a new membrane $[ \; ]_1$ enclosing $SI$ and all *bin membranes*:

     $[ \; [ \; \cdots \; ]SI \; [ \cdots \; ]_{b,1,Bi,I}^0$ , $i \in \{1 ..n\}$ $]_1$

   - For Rules 3 -6: *repeat* the following *until Termination condition* is true:

     - Check applicability.

     - Select randomly a rule for execution.

     - Check termination condition.

   - *Rule 3*: $[\ldots w_i]_{SI}^k$ , $\rightarrow \; w_{i,c} \; [\ldots]_{SI}^{k-1}$ , $k \geq 1$ , $c \in \{FF, BF, WF\}$.

     /* This out-communication rule sends out (of the subset membrane) an item. The item evolves to indicate the selection criterion used in its future packing. The three criterions are used with equal probabilities. */

   - *Rule 4*: $w_{i,c} \; [\ldots]_{b,x,Bi,I}^k \rightarrow [\ldots w_i]_{b,x,Bi,I}^{k+w_i}$ , $Bi > k \geq 0$ , $c \in \{FF, BF, WF\}$.

     /* This in-communication rule sends out an item. $Bi$ is determined using the selection criterion $c$. */

- *Rule 5*: $[\ldots]_{b,x,Bi,I}^{Y} \rightarrow [\ldots]_{b,x,Bi,I}^{Y,-} \ [\ldots]_{b,x+1,Bi,I}^{0} \ , Y \geq Bi \ / \ 2.$

  /* This division rule creates a new bin of the same type. The change of polarity prevents the rule to be executed more than once for the same bin. */

- *Rule 6*: $[\ldots]_{SI}^{0} \rightarrow Yes.$ // All items are sent out of *SI* membrane.

  /* This rule allows for detecting the end of thread computation (Termination condition). A simple *atomic* counter can be checked to make sure that all items are packed. Otherwise, the execution loop continues allowing for Rule 4 to pack remaining items in the membrane. */

5. All threads within a block will synchronize (*Syncthreads()* ) before return.
6. The host executes a call to (*cudaDeviceSynchronize()* ) to make sure that the kernel has terminated then the results can be copied from device memory to host.

**Remarks on correctness:**

1. No synchronization is required in rule execution since: 1- items list is partitioned between threads (in the first heuristic) or blocks (in the second heuristic), 2- each thread is executing one of the applicable rules at a time. (The parallel execution is implemented by CUDA blocks and threads.)
2. An item can be packed in a bin only if there is a sufficient space (Rule 4).
3. A new bin is generated for each type only if all bins of the same type are used to at least half of their capacity (Rule 5). Therefore, the rule guarantees that only a finite number of bins of each type will be generated in the system.

## 6.4. Second Heuristic

The second heuristic differs from the first heuristics in that:

- Distribute the items into subsets of size *s*, such that *s! ≤ Maximum block size*.
- Find all the permutations of each subset in parallel and assign each permutation to a thread within a block.
- Each thread *packs* (*Pack_2*) the items in its subset according to their order in the permutation using rules 2-6.
- After all thread items are packed, thread 0 of each block computes the minimum capacity used by any thread in its block, and the index of the corresponding thread.
- The best results are accumulated by the host.

## 7. NUMERICAL RESULTS

In this experiment we have used Intel(R) Core (TM) i5-2410M CPU @ 2.30 GHz, 6GB RAM, GeForce GTX 560M, Driver 285.86, CUDA Cores 192, and CUDA Toolkit CUDA 4.1.28. We use the benchmarks that were used by (Coreia, I., Goveia, L. and da Gamma, F., 2008). They generate two classes of benchmarks. In this experiment, we used the first class. For some experiments we need large number of items which are not available in the tested benchmarks; for that we created new instances such that:

1. Items weights are selected randomly from the set {1, 2, …, 20}
2. Number of items (***m***) ∈ {3, 5, 10, 15, 30, 5000, 10000, 50000, 100000}

We grouped the whole instances which are used in our experiments into the following groups:

**Group 1:** consists of nine instances where: ***m*** ∈ {3, 5, 10, 15, 30, 100, 200, 500, 1000}, ***b*** = 3, types of bins = {100, 200, 300}

**Group 2:** consists of five instances, where the weight of the items is selected manually: ***m*** = 1000, ***b*** = 4, types of bins = {10, 20, 30, 40}, the weights of the items and the number of the existence of the item in each instance are:

a. {2(250), 5(150), 11(150), 14(150), 15(150), 20(150)}
b. {7(200), 10(200), 12(200), 14(200), 15(200)}
c. {2(100), 4(100), 5(100), 8(100), 9(100), 13(100), 15(100), 16(100), 18(100), 20(100)}
d. {1(150), 3(50), 4(100), 5(50), 7(100), 9(50), 10(50), 11(50), 12(50), 13(100), 15(50), 16(50), 18(100), 20(50)}
e. {3(500), 11(500)}

**Group 3:** consists of eight instances: ***m*** ∈ {100, 200, 500, 1000, 5000, 10000, 50000, 100000}, ***b*** = 3, types of bins = {100, 200, 300}

Where, the instances with number of ***m*** ∈ {100, 200, 500, 1000} are taken from (Coreia, I., Goveia, L. and da Gamma, F., 2008) the lists of items number: (100_1, 200_1, 500_1, 1000_1). Meanwhile, the instances with number of ***m*** ∈ {5000, 10000, 50000, 100000} are generated randomly.

**Group 4:** consists of twelve instances: These instances are the instances of (Coreia, I., Goveia, L. and da Gamma, F., 2008) which are mentioned at the beginning of this section.

It is important to know that the computation capability of the used machine is limited. For that, we had to divide some problem into parts (sub-kernels). A separate kernel will execute each part. At the end, we combine the results of the kernels to get the final result. For example for 100000 items and heuristic 2, we use 625 kernels each with 32 blocks, and each Block size will not exceed 120 threads. This size is very small and affects the performance of the heuristics, but it is suitable for our machine and guarantees that the system will work without any unexpected memory over flow or device timeout.

Two types of kernels are used. The first kernel is for packing the items and the second kernel is for computing the total capacity of the used bins. The notation $1^{st}K$ stands for the execution time of the first kernel. Where, the notation $2^{nd}K$ stands for the execution time of the second kernel. In some cases there is only one block of threads, which mean the total capacity of the used bins is computed in the first kernel, and the execution time of $2^{nd} K$ will be equal to 0. All the times that were reported in our experiments are GPU time, except in serial implementation it is the CPU time.

## Comparison between the Two P System Heuristics

We use Group 3 of instances. In Table 1 we compare the execution structure of our two heuristics. Grid size and the number of the kernels depend on the total number of the items $m$ and the used heuristic. It is clear in Table 1 that the first heuristic needs less number of blocks and kernels which implies a faster performance as shown in Table 2. Table 3 shows the best solution that was reached by the two heuristics. We notice that the first heuristic is able to bring better solution than the second heuristic because it deals with less number of blocks and the number of items that was assigned to a thread is larger than the second heuristic.

**Table 1.** The execution structure of the two heuristics

| M | First Heuristic | | | | Second Heuristic | | | |
|---|---|---|---|---|---|---|---|---|
| | Grid size | | Items/ thread | Number of Kernels | Grid size | | Items/ block | Number of Kernels |
| | Block | Thread | | | block | thread | | |
| 100 | 1 | 10 | 10 | 1 | 20 | 120 | 5 | 1 |
| 200 | 1 | 20 | 10 | 1 | 40 | 120 | 5 | 2 |
| 500 | 1 | 50 | 10 | 1 | 100 | 120 | 5 | 4 |
| 1000 | 1 | 100 | 10 | 1 | 200 | 120 | 5 | 7 |
| 5000 | 1 | 500 | 10 | 1 | 1000 | 120 | 5 | 32 |
| 10000 | 1 | 1000 | 10 | 1 | 2000 | 120 | 5 | 63 |
| 50000 | 5 | 1000 | 10 | 1 | 10000 | 120 | 5 | 313 |
| 100000 | 10 | 1000 | 10 | 1 | 20000 | 120 | 5 | 625 |

**Table 2.** The execution time of the two heuristics using instances of Group 3

| M | First Heuristic | Second Heuristic | |
|---|---|---|---|
| 100 | 0.017623 | 1st K: 0.019020 | 2nd K: 0.08876 |
| 200 | 0.017679 | 1st K: 0.023359 | 2nd K: 0.09188 |
| 500 | 0.017626 | 1st K: 0.040346 | 2nd K: 0.01241 |
| 1000 | 0.018638 | 1st K: 0.094445 | 2nd K: 0.01151 |
| 5000 | 0.019600 | 1st K: 0.376696 | 2nd K: 0.012366 |
| 10000 | 0.019165 | 1st K: 0.765959 | 2nd K: 0.024270 |
| 50000 | 1st K: 0.013015   2nd K: 0.013174 | 1st K: 3.575214 | 2nd K: 0.111391 |
| 100000 | 1st K: 0.014671   2nd K: 0.011624 | 1st K: 5.782224 | 2nd K: 0.202711 |

These results show the efficiency of the first heuristic, and its speed comparing to the second heuristic. Also it shows the ability of the first heuristic to bring same or better results comparing to the second heuristic with less time and space for when using small instances.

**Table 3.** The results of the two heuristics in case of using instances of Group 3

| m | Total weight of items | First Heuristic | Second Heuristic |
|---|---|---|---|
| | | Best Solution Reached | Best Solution Reached |
| 100 | 1119 | 1,600 | 2,300 |
| 200 | 2171 | 3,800 | 4,300 |
| 500 | 5511 | 9,900 | 10,600 |
| 1000 | 10459 | 21,400 | 21,200 |
| 5000 | 53252 | 117,100 | 112,600 |
| 10000 | 105516 | 226,600 | 200,800 |
| 50000 | 526,327 | 1,153,900 | 1,009,200 |
| 100000 | 1,051,854 | 2,308,800 | 2,016,200 |

## Comparison with a Serial and Parallel Implementations

To know the effectiveness of the parallelism of our P system, we compare its CUDA implementation with a serial and parallel implementation. For this purpose, we implement a serial application that computes exact solution by computing all the permutation of a list of up to 10 items. We then pack the each permutation inside bins with different sizes. Finally, we find the permutation that yield the minimum capacity of the used bins. Also, we implemented another CUDA application to find the exact solution of the VSBPP by packing the permutations in parallel using BF, WF, and FF. Then find the permutation that uses the least capacity of bins. In this implementation, we assigned each permutation to a thread. Then within a block find the permutation that used the least capacity of bins. Then, within a grid find the permutation that used the least capacity of bins which will be the final results. We used the instances in Group 1. Table 4 shows the execution time and the number of the permutations that have been processed by our second heuristic and the serial implementation. Table 5 lists the processing time of the second heuristic and the parallel implementation And Table 6 shows the execution structure of our second heuristic and the parallel implementation.

**Table 4.** The execution time required by the Second heuristic and the Serial Implementation

| M | Second Heuristic | | Serial Implementation | |
|---|---|---|---|---|
| | Permutations | Time (sec.) | Permutations | Time (sec.) |
| 3 | 6 | $1^{st}$ K: 0.011561 | 6 | 0.009386 |
| 5 | 120 | $1^{st}$ K: 0.013344 | 120 | 0.282215 |
| 10 | 240 | $1^{st}$ K: 0.016758  $2^{nd}$ K: 0.01098 | 3628800 | 592.81012 |
| 15 | 360 | $1^{st}$ K: 0.017598  $2^{nd}$ K: 0.01091 | NA | NA |
| 30 | 720 | $1^{st}$ K: 0.018640  $2^{nd}$ K: 0.01112 | NA | NA |
| 100 | 2400 | $1^{st}$ K: 0.019020  $2^{nd}$ K: 0.08876 | NA | NA |
| 200 | 4800 | $1^{st}$ K: 0.023359  $2^{nd}$ K: 0.09188 | NA | NA |
| 500 | 12000 | $1^{st}$ K: 0.040346  $2^{nd}$ K: 0.01241 | NA | NA |
| 1000 | 24000 | $1^{st}$ K: 0.094445  $2^{nd}$ K: 0.01151 | NA | NA |

It is clear from these results the second heuristic is faster than the Serial and the Parallel Implementation. Also, the increase of the processing time of the second heuristic is steady and slow unlike the Serial and the Parallel Implementation. Also, we can notice that the Parallel Implementation can find a solution for VSBPP in a reasonable time for a list with at most 10 items. However, the second heuristic is the best in achieving the minimum processing time for large instances.

**Table 5.** The execution time of the second heuristic and the parallel implementation

| M | Second Heuristic | | All permutations parallel Implementation |
|---|---|---|---|
| | Time of All Kernels (sec.) | | Time of All Kernels (sec.) |
| 3 | $1^{st}$ K: 0.011561 | $2^{nd}$ K: 0.0 | 0.011293 |
| 5 | $1^{st}$ K: 0.013344 | $2^{nd}$ K: 0.0 | 0.016814 |

| 10 | 1st K: 0.016758 | 2nd K: 0.01098 | 1.022951 |
| 15 | 1st K: 0.017598 | 2nd K: 0.01091 | 15.677255 *100 kernels only* |
| 30 | 1st K: 0.018640 | 2nd K: 0.01112 | NA |
| 100 | 1st K: 0.019020 | 2nd K: 0.08876 | NA |
| 200 | 1st K: 0.023359 | 2nd K: 0.09188 | NA |
| 500 | 1st K: 0.040346 | 2nd K: 0.01241 | NA |
| 1000 | 1st K: 0.094445 | 2nd K: 0.01151 | NA |

**Table 6.** The execution structure of the second heuristic and the Parallel Implementation

| M | Second Heuristic | | | | All Permutations Parallel Implementation | | | |
| | Grid size | | Items/ block | Number of Kernels | Grid size | | Items/Th read | Number of Kernels |
| | block | thread | | | block | thread | | |
| 3 | 1 | 6 | 3 | 1 | 1 | 6 | 3 | 1 |
| 5 | 1 | 120 | 5 | 1 | 1 | 120 | 5 | 1 |
| 10 | 2 | 120 | 5 | 1 | 5184 | 70 | 10 | 10 |
| 15 | 3 | 120 | 5 | 1 | 5184 | 70 | 15 | 3603600 |
| 30 | 6 | 120 | 5 | 1 | NA | NA | NA | NA |
| 100 | 20 | 120 | 5 | 1 | NA | NA | NA | NA |
| 200 | 40 | 120 | 5 | 2 | NA | NA | NA | NA |
| 500 | 100 | 120 | 5 | 4 | NA | NA | NA | NA |
| 1000 | 200 | 120 | 5 | 7 | NA | NA | NA | NA |

## The Packing Utilization of The second heuristic

To measure the efficiency of a heuristic for the VSBPP, two criteria are required, the execution time and the packing utilization. The packing utilization is computed using the formula:

$$U = \frac{\text{Total weight of items}}{\text{best Solution Reached}}$$

If the achieved utilization of a packing is large that means that the waste space in the used bin is minimized, which is an indication of the robustness of the tested heuristic (Ortmann, F., Ntene, N., and Van Vuuren, J., 2010). Table 7 shows the best results that achieved when packing the instances of Group 2 using the second heuristic. The packing utilization is about 89% of the capacity of the used bins. This is an indication that our heuristic is efficient and robust.

**Table 7.** The packing utilization of the second heuristic using instances of Group 2

| sequence of Instance | Total Items Weight | Best Reached Solution | Utilization |
|---|---|---|---|
| 1 | 10,250 | 11,490 | 0.892 |
| 2 | 11,600 | 12,990 | 0.892 |
| 3 | 11,000 | 12,260 | 0.897 |
| 4 | 9,400 | 10,460 | 0.898 |
| 5 | 7,000 | 8,310 | 0.842 |

### Comparison with Discretized Formulations

To ensure the robustness and the efficiency of our P system heuristics we need to use the instances of Group 4 which are the instances of (Coreia, I., Goveia, L. and da Gamma, F., 2008). In their paper, they introduced results of number of Discretized Formulations. They used CPLEX solver to implement their formulations. In Table 8, we compare the execution time of 3 of their Discretized Formulations and our second heuristic, in case of packing 1000 items using (3, 6, 12) types of bins.

**Table 8.** The execution time of the Second heuristic and three discretized formulations in case of using 3, 6 and 12 types of Bins

| Models | 3 Bins Types (sec.) | 6 Bins Types (sec.) | 12 Bins Types (sec.) |
|---|---|---|---|
| Second Heuristic | 1$^{st}$ K: 0.094445 2$^{nd}$ K: 0.011513 | 1$^{st}$ K: 0.098116 2$^{nd}$ K: 0.010878 | 1$^{st}$ K: 0.104158 2$^{nd}$ K: 0.011004 |
| P1+(6)+(7) | 576 | 1611 | 2885 |
| P2+(16)+(17)+(19)+(20) | 1614 | 3323 | 5419 |
| P2+(13)+(14)+(16)+(17)+(19)+(20) | - | - | - |

It's clear that the processing time of the second heuristic is much less than the processing time of the three discretized formulations. Also we notice that in our implementation there is a small difference between the execution time in case of using 3, 6, or 12 bins types. This is because in our CUDA implementation we rely on the local and shared variables and simple arrays structures. Meanwhile in the three discretized formulations there was a big difference between the execution time in case of using 3, 6, or 12 different bins types, which mean another advantage added to our P system heuristic.

In Table 9 we are listing the results of the instances in Group 4. Dr. Isabel Coreia provided us with (commendable) their detailed results. According to their paper *D* is a constraint representing "the minimum number of bins of each size that must be considered in order to have a total available capacity greater than or equal than the total requirement" (Coreia, I., Goveia, L. and da Gamma, F., 2008). In our experiment it is not possible to put this constraint because of the nature of our P system and our CUDA implementation. In our P system we are dealing with sublists of the main list which could use number of bins less than the number of bins types. For that we list in table 9 the best reached solution, such that it is using the maximum number of bins types.

**Table 9.** The results of the second heuristic compared to the optimum results of Coreia Paper in case of using 3, 6, and 12 types of Bins

| instance | Optimum 3 types of bins | | Second heuristic 3 types of bins | Optimum 6 types of bins | | Second heuristic 6 types of bins | Optimum 12 types of bins | | Second heuristic 12 types of bins |
|---|---|---|---|---|---|---|---|---|---|
| | D=2 | D=3 | | D=2 | D=3 | | D=1 | D=2 | |
| 100_1 | 6292 | 6196 | 2,300 | 6269 | 6196 | 1,550 | 6293 | 6240 | 1,375 |
| 100_2 | 8292 | 7610 | 2,300 | 7333 | 6777 | 1,700 | 7337 | 6703 | 1,525 |
| 100_3 | 8292 | 7610 | 2,300 | 7333 | 6777 | 1,700 | 7337 | 6703 | 1,500 |
| 100_4 | 8292 | 7610 | 2,300 | 7333 | 6777 | 1,750 | 7471 | 6780 | 1,550 |
| 100_5 | 8292 | 7610 | 2,300 | 7333 | 6777 | 1,700 | 7337 | 6703 | 1,475 |
| | D=4 | D=5 | | D=3 | D=4 | | D=2 | D=3 | |
| 200_1 | 14584 | 13902 | 4,300 | 13991 | 13252 | 3,250 | 13808 | 13165 | 2,850 |
| 200_2 | 14584 | 13902 | 4,300 | 13991 | 13252 | 3,200 | 13808 | 13165 | 2,950 |
| 200_3 | 14584 | 13902 | 4,300 | 13991 | 13252 | 3,250 | 13808 | 13165 | 2,975 |
| 200_4 | 13584 | 12902 | 4,300 | 12767 | 12671 | 3,050 | 12666 | 12458 | 2,625 |
| 200_5 | 16584 | 15316 | 4,300 | 15181 | 14476 | 3,400 | 14856 | 14332 | 3,125 |
| | D=10 | D=11 | | D=6 | D=7 | | D=3 | D=4 | |
| 500_1 | 32460 | 31778 | 10,600 | 33258 | 32382 | 7,500 | 34529 | 32406 | 6,950 |
| 500_2 | 37460 | 35606 | 11,000 | 37413 | 35537 | 8,250 | 39220 | 35506 | 7,500 |
| 500_3 | 34460 | 33192 | 10,600 | 35189 | 33796 | 7,950 | 36647 | 33980 | 7,350 |
| 500_4 | 36460 | 34606 | 10,600 | 36706 | 35313 | 8,250 | 38513 | 35302 | 7,950 |
| 500_5 | 35460 | 34192 | 10,900 | 35482 | 34089 | 7,950 | 37147 | 34086 | 7,325 |
| | D=18 | D=19 | | D=11 | D=12 | | D=6 | D=7 | |
| 1000_1 | 71628 | 69774 | 21,200 | 71168 | 68740 | 15,750 | 72001 | 68621 | 15,476 |
| 1000_2 | 71628 | 69774 | 21,700 | 71461 | 68964 | 15,750 | 72160 | 69345 | 15,600 |
| 1000_3 | 72628 | 70774 | 21,500 | 72461 | 69964 | 15,950 | 73160 | 69345 | 15,425 |
| 1000_4 | 72628 | 70774 | 21,600 | 72168 | 69671 | 15,800 | 72794 | 69121 | 15,050 |
| 1000_5 | 72628 | 70774 | 21,700 | 72168 | 69671 | 15,650 | 72794 | 69121 | 15,725 |

## 8. CONCLUSION AND FUTURE WORK

Finding minimum space or cost to pack a certain list of items is a critical problem to many applications; the VSBPP is a variant of this problem. The VSBPP is one of the classical combinatorial NP-Hard Problems. In the literature of the VSBPP, there are few parallel heuristics that were presented to find a solution for some instances of this Problem. In this paper, we propose two parallel heuristics that use the approach of the membrane computing to find a solution for the VSBPP using NVIDIA's CUDA. Membrane computational model is parallel, distributed, scalable and nondeterministic. These features makes it very suitable model to use in GPGPU and CUDA. In this paper we propose a new variation of membrane computing that we call a *Hybrid P system*. Our model allows for using objects and membrane label and polarity to represent properties instead of using object multiplicity as in the classical model. This new model improves significantly the time and space complexity of implementing the classical P-system.

Based on the numerical results of the CUDA simulation of our two heuristics, the first heuristic is faster than the second one in all cases. But the second heuristic succeeds in achieving better results in case of using benchmarks with large number of items. The results also show that the performance of the second heuristic is surpassing the performance of the serial implementation and the performance of all permutations parallel implantation. The average utilization achieved by the second heuristic is about 89% of the capacity of used bins, which indicates that it is an efficient and robust heuristic. The execution time required by the second heuristic is much less than the time required by the discretized formulations (Coreia, I., Goveia, L. and da Gamma, F., 2008). Also the total capacity of the used bins achieved by the heuristics is less than the best results obtained by these discretized formulations.

The membrane computational model is a good framework for parallel heuristics since it is making the non-determinism and parallelization process easy to implement. The exploitation of the power of the NVIDIA's GPU helped us in getting good results and strengthened the success of the membrane computing. The merge of Membrane Computing model and NVIDIA's CUDA is expected to lead to good results for most of the NP-Complete problems.

In future, we are planning to enhance our Recognizer P system with Active Membrane of the VSBPP in order to increase the packing utilization. We will use the features of the NVIDIA's Kepler to improve the CUDA implementation. Also, we will find a way to make a general reusable P system so that it can be used to solve different kinds of problems.